\documentclass[article,12pt,preprint]{aastex}
\usepackage[margin=1in]{geometry}
\usepackage{graphicx}
\usepackage{epsfig}
\usepackage{fullpage}
\usepackage{keyval}
\title{Discovery and Early Multi-Wavelength Measurements of the Energetic Type Ic Supernova PTF12gzk: A Massive-Star Explosion in a Dwarf Host Galaxy}

\begin{document}

\author{Sagi Ben-Ami\altaffilmark{1,2}, Avishay
  Gal-Yam\altaffilmark{1}, Alexei V. Filippenko\altaffilmark{3}, Paolo
  A. Mazzali\altaffilmark{4,5,6}, Maryam Modjaz\altaffilmark{7}, Ofer
  Yaron\altaffilmark{1}, Iair Arcavi\altaffilmark{1}, S. Bradley Cenko\altaffilmark{3}, 
  Assaf Horesh\altaffilmark{8}, D. Andrew Howell\altaffilmark{9,10},
  Melissa L. Graham\altaffilmark{9,10}, J. Chuck Horst\altaffilmark{11},
  Myunshin Im\altaffilmark{12}, Yiseul Jeon\altaffilmark{12}, 
  Shrinivas R. Kulkarni\altaffilmark{8}, Douglas C. Leonard\altaffilmark{11}, 
  Daniel Perley\altaffilmark{8}, Elena Pian\altaffilmark{13,14,6}, David J. Sand\altaffilmark{9,10,15},  
  Mark Sullivan\altaffilmark{16}, Juliette C. Becker\altaffilmark{8}, 
  David Bersier\altaffilmark{17}, Joshua S. Bloom\altaffilmark{3,18}, 
  Michael Bottom\altaffilmark{8}, Peter J. Brown\altaffilmark{19}, 
  Kelsey I. Clubb\altaffilmark{3}, Ben Dilday\altaffilmark{13,14}, 
  Richard C. Dixon\altaffilmark{20}, Aryeh L. Fortinsky\altaffilmark{1}, 
  Derek B. Fox\altaffilmark{21}, Luis A. Gonzalez\altaffilmark{22}, Avet Harutyunyan\altaffilmark{23}, 
  Mansi M. Kasliwal\altaffilmark{24}, Weidong Li\altaffilmark{3,25},
  Matthew A. Malkan\altaffilmark{26}, Ilan Manulis\altaffilmark{1},
  Thomas Matheson\altaffilmark{27}, Nicholas
  A. Moskovitz\altaffilmark{28}, Philip S. Muirhead\altaffilmark{8},
  Peter E. Nugent\altaffilmark{3,17}, Eran O. Ofek\altaffilmark{1}, 
  Robert M. Quimby\altaffilmark{29}, Joseph W. Richards\altaffilmark{3,30},
  Nathaniel R. Ross\altaffilmark{26}, Kinchen J. Searcy\altaffilmark{31}, 
  Jeffrey M. Silverman\altaffilmark{3}, Nathan Smith\altaffilmark{32}, 
  Andrew Vanderburg\altaffilmark{3}, and Emma S. Walker\altaffilmark{13}
}

\altaffiltext{1}{Department of Particle Physics and Astrophysics, The Weizmann Institute of Science, Rehovot 76100, Israel.}
\altaffiltext{2}{email:sagi.ben-ami@weizmann.ac.il .}
\altaffiltext{3}{Department of Astronomy, University of California, Berkeley, CA 94720-3411, USA.}
\altaffiltext{4}{INAF, Osservatorio Astronomico di Padova, Italy.}
\altaffiltext{5}{Max-Planck Institute for Astrophysics, Garching, Germany.}
\altaffiltext{6}{Kavli Institute for Theoretical Physics, Kohn Hall, University of California, Santa Barbara, CA 93106-4030, USA.}
\altaffiltext{7}{New York University, Center for Cosmology and Particle Physics, Department of Physics, 4 Washington Place, New York, NY 10003, USA.}
\altaffiltext{8}{Cahill Center for Astrophysics, California Institute of Technology, Pasadena, CA, 91125, USA.}
\altaffiltext{9}{Las Cumbres Observatory Global Telescope Network, 6740 Cortona Drive, Suite 102, Santa Barbara, CA 93117, USA.} 
\altaffiltext{10}{Department of Physics, Broida Hall, University of California, Santa Barbara, CA 93106, USA.} 
\altaffiltext{11}{Department of Astronomy, San Diego State University, San Diego, CA 92182, USA.}
\altaffiltext{12}{CEOU/Astronomy Program, Dept. of Physics \& Astronomy, Seoul National University, Seoul, Korea.}
\altaffiltext{13}{Scuola Normale Superiore di Pisa, Piazza dei Cavalieri 7, 56126 Pisa, Italy.}
\altaffiltext{14}{INAF, Astronomical Observatory of Trieste, Via G.B. Tiepolo 11, I-34143 Trieste, Italy.}
\altaffiltext{15}{Harvard Center for Astrophysics and Las Cumbres Observatory Global Telescope Network Fellow.}
\altaffiltext{16}{Department of Physics (Astrophysics), University of Oxford, DWB, Keble Road, Oxford, OX1 3RH, UK.}
\altaffiltext{17}{Astrophysics Research Institute, Liverpool John Moores University, UK.}
\altaffiltext{18}{Physics Division, Lawrence Berkeley National Laboratory, Berkeley, CA 94720, USA.}
\altaffiltext{19}{George P. and Cynthia Woods Mitchell Institute for Fundamental Physics \& Astronomy, Department of Physics and Astronomy, Texas A\&M University, 4242 AMU, College Station, TX 77843, USA.}            
\altaffiltext{20}{Department of Physics \& Engineering, Palomar College, San Marcos, CA 92069, USA.}
\altaffiltext{21}{Department of Astronomy \& Astrophysics, 525 Davey Laboratory, Penn State University, University Park, PA 16802, USA.}
\altaffiltext{22}{Department of Physics, University of California, San Diego, La Jolla, CA 92093, USA.}
\altaffiltext{23}{Galileo Galilei-INAF, TNG, Tenerife, Spain.}
\altaffiltext{24}{Observatories of the Carnegie Institution for Science, 813 Santa Barbara St, Pasadena, CA 91101, USA.}
\altaffiltext{25}{Deceased 2011 December 12.}
\altaffiltext{26}{Department of Physics and Astronomy, University of California, Los Angeles, CA 90095, USA.}
\altaffiltext{27}{National Optical Astronomy Observatory, NOAO System Science Center, 950 North Cherry Avenue, Tucson, AZ 85719, USA.}
\altaffiltext{28}{Carnegie Institution for Science, Department of Terrestrial Magnetism, 5241 Broad Branch Road, Washington, DC 20008 USA.}
\altaffiltext{29}{Kavli IPMU, University of Tokyo, 5-1-5 Kashiwanoha, Kashiwa-shi, Chiba, 277-8583, Japan.}
\altaffiltext{30}{Department of Statistics, University of California, Berkeley, CA 94720-7450, USA.}
\altaffiltext{31}{San Diego Astronomy Association, P.O. Box 23215, San Diego, CA 92193-33125, USA.}
\altaffiltext{32}{Steward Observatory, University of Arizona, 933 North Cherry Avenue, Tucson, AZ 85721, USA.}

\begin{abstract}
We present the discovery and extensive early-time observations of the
Type Ic supernova (SN) PTF12gzk. Our light curves show
a rise of 0.8\,mag within 2.5\,hr. Power-law fits [$f(t)\propto(t-t_0)^n$] 
to these data constrain the explosion date to within one day.  
We cannot rule out a quadratic fireball model, but 
higher values of $n$ are possible as well for larger areas in the fit parameter space.
Our bolometric light curve and a dense spectral
sequence are used to estimate the physical parameters of the exploding
star and of the explosion.  We show that the photometric evolution of
PTF12gzk is slower than that of most SNe~Ic. The high ejecta expansion
velocities we measure ($\sim30,000$\,km\,s$^{-1}$ derived from line minima four days after explosion)
are similar to the observed velocities of broad-lined SNe~Ic 
associated with gamma-ray bursts
(GRBs) rather than to normal SN~Ic velocities. 
Yet, this SN does not show the persistent broad lines 
that are typical of broad-lined SNe~Ic.
The host-galaxy characteristics are also consistent with
GRB-SN hosts, and not with normal SN~Ic hosts. 
By comparison with the spectroscopically similar SN~2004aw, we
suggest that the observed properties of PTF12gzk indicate an initial
progenitor mass of 25--35\,M$_\odot$ and a large [(5--10) $\times
10^{51}$\,erg] kinetic energy, the later being close to the regime of GRB-SN
properties.  
\end{abstract}

\section{Introduction}

A core-collapse supernova (CCSN) occurs when a star having an initial
mass $M\geq8\,{\rm M}_\odot$ ends its life in a catastrophic explosion.
Observationally, CCSNe are divided
into three groups based on their observed spectra: SNe~II show
large amounts of hydrogen, SNe~Ib exhibit helium but little or no
hydrogen, and SNe~Ic do not show significant amounts of hydrogen or
helium \citep[for a review, see][]{Filippenko1997}.

SNe~Ic are heterogeneous. Their luminosity, ejected mass, and kinetic energy span over an order of
magnitude, from the subluminous SN 2004aw to the overluminous SN
1998bw \citep{Mazzali2009,Drout2011}. The light-curve shapes of different events
are also quite diverse. A subclass of SNe~Ic whose spectra are
characterized by broad lines (Type Ic-BL; prototype SN 1998bw) is the
only one for which clear evidence of an association with gamma-ray
bursts (GRBs) exists (GRB-SNe; see Woosley \& Bloom 2006 for a
review). Superluminous SNe (SLSNe) of Type Ic are even more powerful
\citep[][and references therein]{GalYam2012}, but these probably
result from a different physical mechanism. While SNe~Ic are common in
the center of high-metallicity galaxies \citep{Anderson2012}, SLSNe-I
and broad-lined GRB-SNe tend to be found in dwarf hosts
\citep[e.g.,][]{Stanek2006,Modjaz2008,Arcavi2010}, giving untargeted sky surveys
an advantage over targeted surveys in detecting these types of cosmic
explosions.

The Palomar Transient Factory (PTF; Law et al. 2009; Rau et al. 2009)
is a wide-field untargeted sky survey which explores the transient
optical sky. It uses the PTF CFH12k camera mounted on the Palomar
48-inch telescope (P48).  PTF's short observing cadence and real-time
capability \citep[e.g.,][]{GalYam2011} enables the discovery and study
of SNe at early stages of the explosion. In this {\it Letter} we report the
discovery and study of PTF12gzk, a peculiar SN~Ic in a dwarf
star-forming galaxy located at redshift $z = 0.0137$ (distance
57.8\,Mpc, distance modulus 33.8\,mag, assuming H$_0$ =
71\,km\,s$^{-1}$\,Mpc$^{-1}$).

\section{Discovery}

PTF12gzk was discovered on 2012 July 24.3 (UT dates are used herein)
at $\alpha$(J2000) = 22$^{\rm h}$12$^{\rm m}$41.53$^{\rm s}$ and
$\delta$(J2000) = $+00^\circ30'43.1''$, in the dwarf galaxy SDSS
J221241.53+003042.7 (within the SDSS Stripe~82 footprint), as part of a
coordinated PTF-JVLA\footnote{The Jansky Very Large Array is operated
  by the National Radio Astronomy Observatory (NRAO), which is a
  facility of the National Science Foundation (NSF), operated under
  cooperative agreement by Associated Universities, Inc. (AUI).}
survey\footnote{It was also independently discovered by the La-Silla
  Quest (LSQ) and Pan-Starrs1 (PS1) surveys, and designated LSQ 2012dwl
  and PS1-12baa, respectively.}.  The discovery magnitude was 20.66 in
the $r$ band\footnote{PTF magnitudes are given in the PTF
  natural-magnitude system \citep{Ofek2012a}, with respect to the SDSS
  $r$-band magnitudes.}, and it was not detected down to mag 21.6
($3\sigma$) in previous PTF images obtained on July 19
\citep{Benami2012}.

Shortly after discovery, we initiated an extensive follow-up campaign,
including our {\it Hubble Space Telescope (HST)}
Target-of-Opportunity (ToO) program for STIS ultraviolet (UV)
spectroscopy of a stripped-envelope SN (Cycle 19, GO-12530; PI
Filippenko) and {\it Swift} X-ray and UV photometry (Cycle 8, PID
8110099; PI Kasliwal), the results of which are presented herein (see
Figures 1 and 2 for photometry and spectroscopy, respectively). We
also triggered radio and millimeter observations using the JVLA
(program 12A-363; PI Horesh) and the Combined Array for Research in
Millimeter-wave Astronomy (program 12A-c0945; PI Horesh); see Horesh
et al. (in preparation).

\section{Observations}
\subsection{Photometry}
Optical photometry of PTF12gzk was obtained using multiple telescopes
(Table 1). All data were calibrated with respect to the SDSS catalog.
Light curves of PTF12gzk are shown in Figure 1. 
\begin{figure}[]
\centering
\scalebox{0.7}{\includegraphics[]{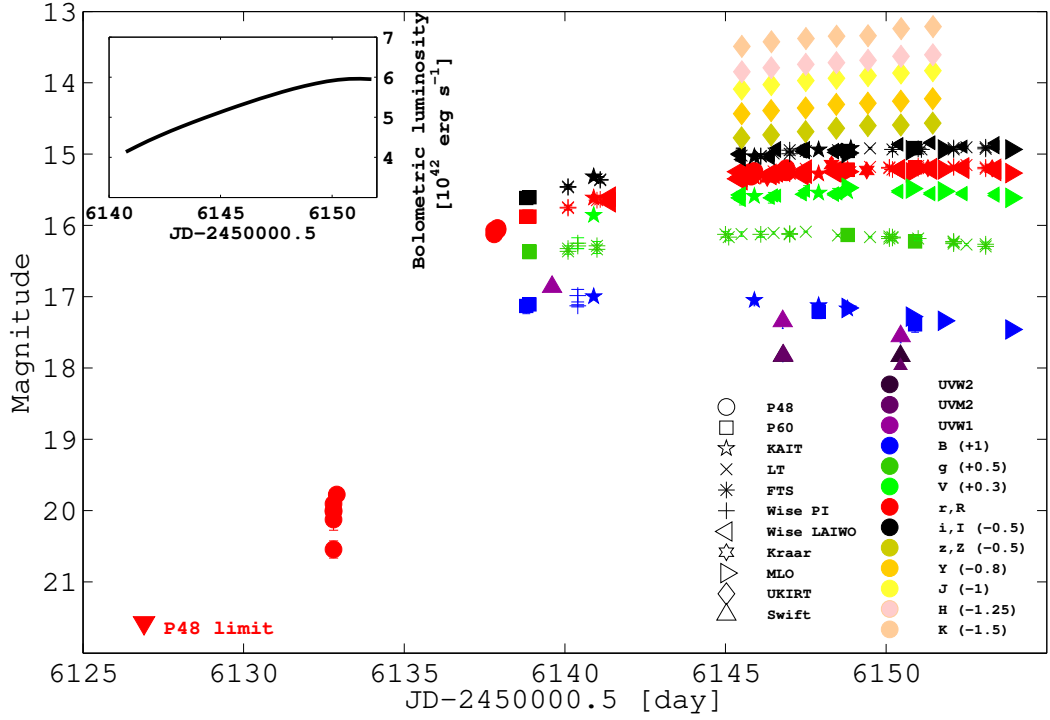}}
\centering
\scalebox{0.9}{{\includegraphics{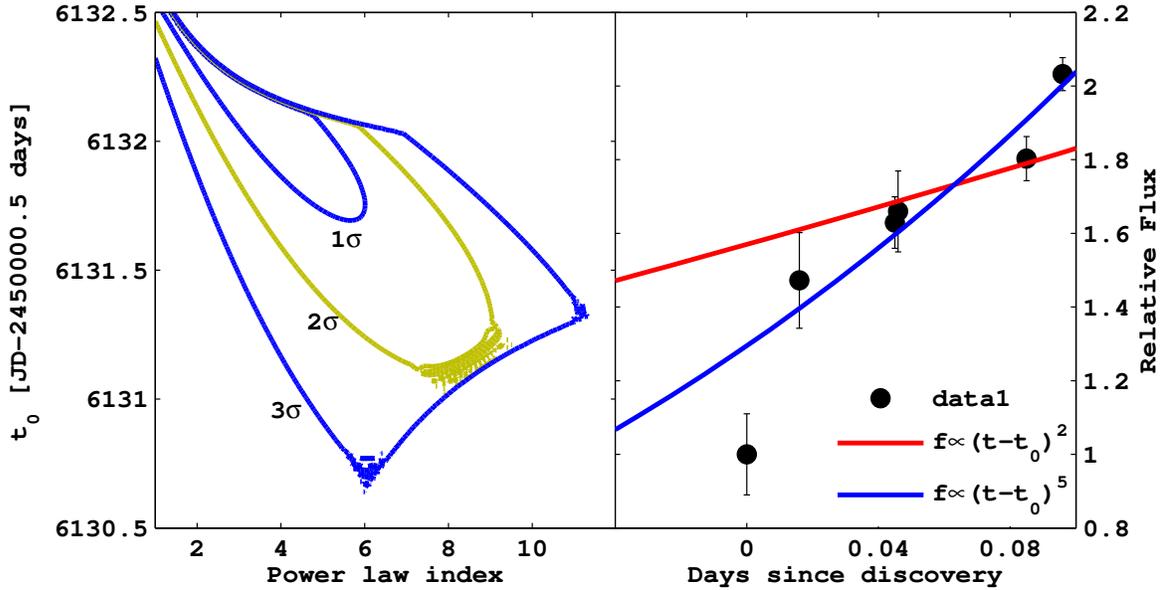}}}
\caption{\scriptsize PTF12gzk photometry. \textbf{Top:} The long rise
  time and the large delay in peak magnitudes between the $B$ and
  $r$ bands is evident in the light curve. \textbf{Inset:} 
  A bolometric light curve derived from a fourth-degree polynomial fit to our
  UV-optical-IR data.
  \textbf{Bottom left:} A $\chi^2$ probability-density-function contour plot comparing power-law models with the first
  2.5\,hr of observations (see text). The value of the index $n$ is not well constrained. Discontinuities in the contour curves are caused
  by finite sampling of the parameter space, and have no statistical significance. 
  \textbf{Bottom right:} Early photometry. The flux is scaled with respect to the first data point. Fits of the scaled flux
  to $f\propto(t-t_0)^2$ (red) and $f\propto(t-t_0)^5$ (blue) are also shown. Considering the flux normalization as an additional 
  parameter, there is no significant difference between the two fits. 
}
\end{figure}
\begin{table}[h!p!t!]
\centering
\begin{tabular}{|c|c|c|c|c|}
  \hline
  Date (MJD)            & Instrument &  Filter & Apparent Magnitude & $1\sigma$ Uncertainty  \\ \hline\hline
2456132.763     &       P48+CFH12k &   $R$       &    20.66             &       0.12  \\
2456132.779     &       P48+CFH12k &   $R$       &    20.24             &       0.15  \\
2456132.808 	&       P48+CFH12k &   $R$ 	   & 	20.13 	          &       0.08  \\
2456132.809     &       P48+CFH12k &   $R$ 	   & 	20.11 	          &       0.12  \\
2456132.848 	&       P48+CFH12k &   $R$ 	   & 	20.02 	          &       0.07  \\  \hline
\end{tabular}
\label{tab:Photometry}
\caption{\scriptsize
PTF12gzk photometry. Data were obtained using the P48+PTF camera, 
the Palomar 60-inch telescope + GRB camera, 
the Fairchild camera on the 40-inch telescope at
Mt. Laguna Observatory (MLO), FS01 on the 2-m Faulkes Telescope South
(FTS), the IO:O camera on the Liverpool telescope (LT), the PI camera
on the Wise 1-m telescope, the Weizmann Institute Kraar 16-inch
telescope, and the 0.76-m Katzman Automatic Imaging Telescope (KAIT;
Filippenko et al. 2001). The full set of photometric data is available
in the electronic version of this paper, as well as from WISeREP,
http://www.weizmann.ac.il/astrophysics/wiserep/; Yaron \& Gal-Yam
  (2012).
}
\end{table}

The optical data were reduced using standard IRAF procedures for
aperture photometry (FTS data were reduced using point-spread-function photometry via
DoPHOT).  We subtract reference templates from
the P48 and P60 data to remove contamination from the host\footnote{SDSS images were used as reference images for the P60 data.}.
Pre-explosion templates were not used for other data, but the
contribution from the underlying galaxy ($g,r,z=19.05$, 19.03, 
18.75\,mag from SDSS) is negligible, as was also demonstrated by reducing P60 data both with and without template subtraction. The data were calibrated to SDSS
stars in the field, using the transformation equations given by Jordi et
al. (2006) to place the local standards on the Johnson-Cousins system\footnote{We tested our cross-calibration among different telescopes by calculating
low-order polynomial fits to the combined dataset in each filter and searching for systematic offsets of particular datasets from the combined fit. We found that
two datasets (KAIT $I$ and LT $g$) had to be corrected by 0.06 mag.}.

Infrared (IR) photometry of PTF12gzk (Table 1) was obtained using the
Wide Field Camera mounted on the United Kingdom Infrared Telescope
(UKIRT-WFCAM) using SExtractor, and calibrated with respect to the 2MASS catalog
(magnitude errors $<0.07$\,mag) using the relation of Hodgkin et
al. (2009).

Based on the absence of strong Na~I~D lines (see section 4), as well as
the blue early-time spectrum (see below), we take a conservative approach assuming negligible 
extinction by the host galaxy at the SN location.

During the first night of observation, PTF12gzk brightened by
$\sim0.8$\,mag in less than 2.5\,hr to 19.85\,mag in $r$; we thus
obtained remarkably early coverage of a SN~Ic. We find that the
SN peaked at $r=15.2$ mag on August 14, $g=15.55$ on August 8, and
$B=16$ on August 4. On August 15, PTF12gzk peaked in the $i$ band.

PTF12gzk was observed with the X-Ray Telescope (XRT) and the
Ultraviolet/Optical Telescope (UVOT) onboard the {\it Swift}
satellite. XRT measurements, beginning at 13:39 on July 31, detected
no source at the location of PTF12gzk; we estimate a dead-time-corrected 
limit on the XRT count rate of $<2\times 10^{-3}$\,cps.
Assuming a power-law spectrum with a photon index of $2$, this
corresponds to a limit on the X-ray flux of $<7\times
10^{-14}$\,erg\,cm$^{-2}$\,s$^{-1}$. 
UVOT data were reduced using a $5''$ aperture to measure the counts for the coincidence-loss correction,  
a $3''$ or $5''$ source aperture (based on the signal-to-noise ratio) for the aperture photometry,              
and applying an aperture correction as appropriate.
We used zeropoints from Breeveld et al. (2011), including a time-dependent sensitivity loss                                             
to put the magnitudes on the UVOT photometric system. Results are given in figure 1.\\ 

We have calculated a bolometric light curve by integrating the flux in
the $UVW2$, $UVM2$, $UVW1$, $B$, $g$, $V$, $r$, $R$, $i$, $I$, $J$, $H$, $K$ filters and fitting a fourth-degree polynomial to the obtained data points. 
When lacking IR photometry, we assume a constant fraction of IR flux, 
found to be $\sim18$\% from synthetic
photometry using TSPEC IR spectra (see \S~4.2). We estimate that the
lack of IR (UV) coverage prior to August 7 introduces an uncertainty
of $\sim10$\% from the small variation seen in the IR contribution
($<5$\% with respect to the overall flux) between August 4 (the first
TSPEC IR spectrum) and August 12.  Uncertainties introduced by the
lack of UV photometry are $\sim5$\% from the even smaller variation in
the UV contribution ($<2$\% relative to the overall flux).  The
bolometric light curve is given in Figure 1 and shows a rise time 
of $18\pm1$ day, similar to that in the $r$ band.

\subsection{Spectroscopy}

Extensive spectroscopy of PTF12gzk was performed, and detailed
analysis will be presented in a future paper.  A selection of optical
spectra is shown in Figure 2.\footnote{All ground-based spectra
  presented in this paper are released on WISeREP,
http://www.weizmann.ac.il/astrophysics/wiserep/; Yaron \& Gal-Yam
  (2012).}

\begin{figure}[]
\centering
\scalebox{1.7}{\includegraphics[width=6cm]{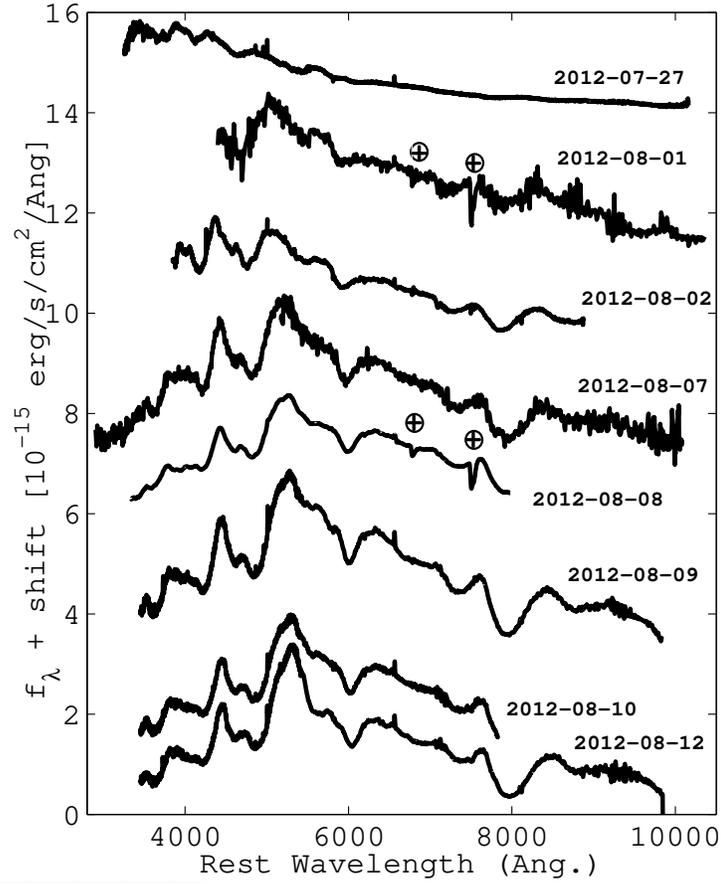}}
\centering
\begin{tabular}{cc}
  \scalebox{0.9}{\raisebox{0.75cm}{\includegraphics{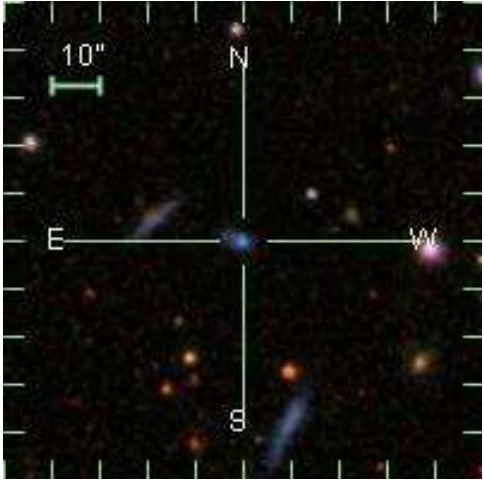}}}&
  \scalebox{0.4}{\includegraphics[angle=90]{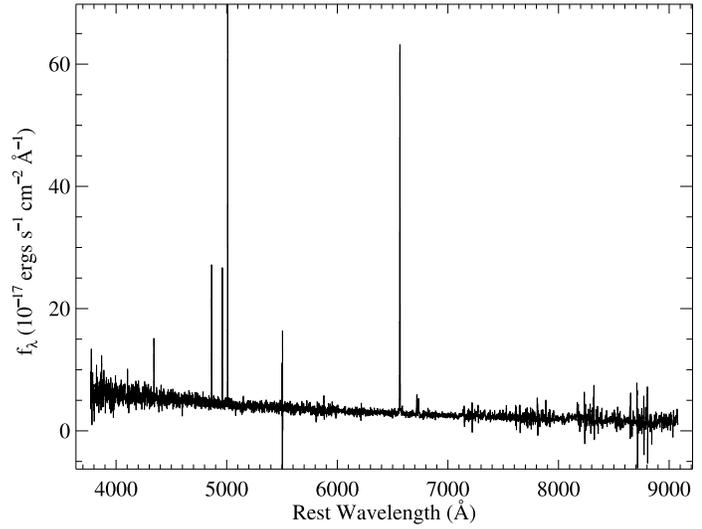}} 
\end{tabular}
\label{fig:IntLines}
\caption{\scriptsize
\textbf{Top:} PTF12gzk optical spectra; telluric features are marked.
\textbf{Bottom}: The host galaxy of PTF12gzk. Left: SDSS
image; note the blue color and small spatial size ($1''=280$\,pc).
Right: SDSS spectrum; note the strong emission lines and blue
continuum.}
\end{figure}     

The first spectrum was obtained on July 27, revealing the object to be
a peculiar SN~I with a blue continuum, reminiscent of the blue
continuum observed in the early-time spectra of SN 2006aj
\citep{Mazzali2006b}. Initially it was difficult to classify the
spectrum securely, but later optical spectra resemble those of SNe~Ic,
with an absence of prominent He~I lines \citep{Benami2012}.

The August 1 spectrum was taken with the FLOYDS spectrograph mounted
on the FTN, a low-dispersion spectrograph ($R\approx400$) with a
single, fixed grating and a cross-dispersing prism, placing the first-
and second-order spectra onto the CCD with a single-exposure
wavelength coverage of 3200--10,000\,\AA.  While the identical twin FLOYDS
spectrographs on FTN and FTS will ultimately be robotically operated,
the present spectrum was taken during hardware commissioning under
human control.

Classification with the help of SNID \citep{Blondin2007} of the August
9 Lick 3-m/Kast spectrum suggests that the best
fit is to the peculiar SN~Ic SN~2004aw near peak brightness
\citep{Taubenberger2006}. A good match is also obtained for PTF10qts, 
a Type Ic-BL SN, at $\sim30\,$days after explosion. 
All spectra show prominent, broad
absorption lines of Ca~II, Si~II, Fe~II, and O~I, with SYNOW fits
yielding maximum velocities of $\sim35,000$\,km\,s$^{-1}$ for the July
27 spectrum, and $\sim20,000$\,km\,s$^{-1}$ for the August 9 spectrum.
See Figure 3 and \S~5 for further discussion.

\begin{figure}[ht]
\centering
\scalebox{1.1}{\includegraphics[width=\textwidth]{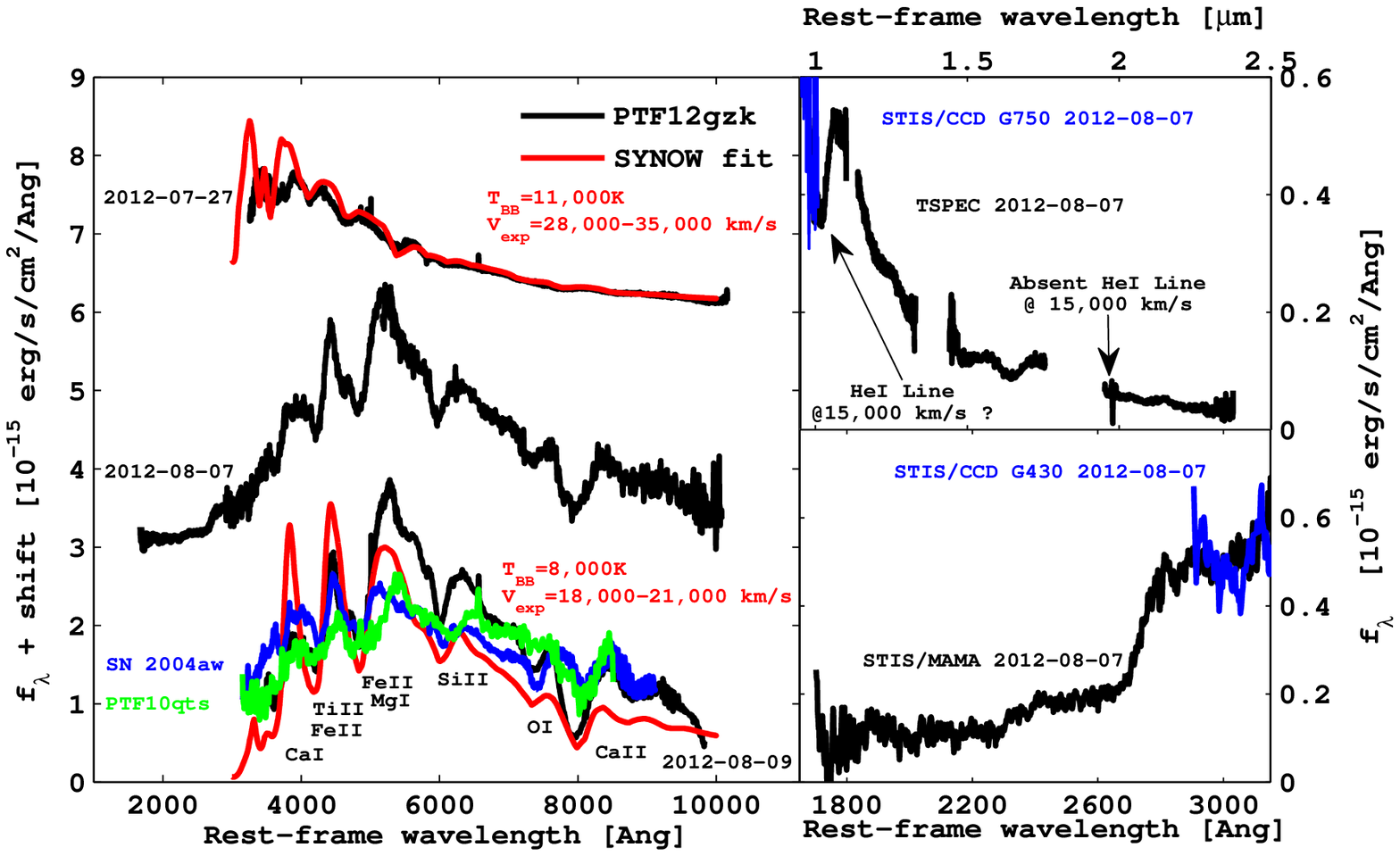}}
\caption{ {\scriptsize \textbf{Left:} Spectroscopic analysis; top
to bottom. The DBSP spectrum taken on July 27. An {\it HST}
STIS/MAMA and STIS/CCD (G430 and G750) spectrum taken on August 7.
A Lick/Kast spectrum taken on August 9. The continuum is
consistent with blackbody temperatures of 11,000\,K and 8000\,K in
the July 27 and August 9 spectra, respectively. A SYNOW fit shows
that the July 27 spectrum is dominated by Ca~II, Ti~II, and Fe~II lines at
velocities of 35,000\,km\,s$^{-1}$, and by O~I, Mg~II, and Si~II at
28,000\,km\,s$^{-1}$; the corresponding velocities for the August 9
spectrum are 21,000 and 18,000\,km\,s$^{-1}$.
The best match suggested by SNID\citep{Blondin2007} 
is to a spectrum of SN 2004aw near peak
brightness (blue curve). A good match is also obtained for PTF10qts, 
a Type Ic-BL SN, at $\sim30\,$days after explosion. 
\textbf{Top right:} The Palomar 5-m TSPEC IR spectrum
obtained on August 7, together with the long-wavelength end of the
{\it HST} STIS/CCD spectrum.
\textbf{Bottom right:} The {\it HST} STIS/MAMA UV spectrum taken 
on August 7. The flux deficit with respect to a blackbody spectrum 
indicates strong line blanketing, evidence for a highly mixed outer
envelope. } }
\end{figure}

The absence of the 2.1\,$\mu$m He~I feature in the IR TSPEC spectrum
taken on August 7 (Figure 3, bottom-right panel) suggests that He is
not abundant in the envelope \citep{Hachinger2012}. While a strong
absorption line is present at 1\,$\mu$m, which could be identified
with He~I 1.0830\,$\mu$m, in this case it is likely contaminated by other
elements such as C, Mg, S, and Ca
\citep[e.g.,][]{Mazzali1998,Sauer2006,Hachinger2012}.

The first {\it HST} UV spectrum, obtained by the Space Telescope
Imaging Spectrometer with the near-UV Multi-Anode Microchannel Array
(STIS/MAMA), was taken between 20:06 August 6 and 03:35 August 7,
with a total exposure time of 11,278\,s. We also obtained optical
spectra with the STIS/CCD, using grisms centered at 4300 and
7500\,\AA.  A mosaic of all three {\it HST} spectra is given in Figure
3, with a focus on the UV spectrum in the bottom-right panel.  The flux
deficit with respect to blackbody spectra seen in the STIS/MAMA UV
spectrum is characteristic of all SNe~I \citep{Panagia2007} and
indicates strong line blanketing, evidence for a highly mixed outer
envelope devoid of hydrogen.  Full analysis of the {\it HST} UV
spectra will be presented by Ben-Ami et al. (in prep.).

\section{Host-Galaxy Analysis}
An image and a spectrum of the host galaxy obtained by the SDSS prior
to explosion are shown in Figure 2. The SN exploded very close to the
center of the host galaxy (offset $0.0\arcsec$ N and $0.4\arcsec$ E).
We downloaded the host-galaxy spectrum from the SDSS 9th Data Release
\citep[DR9;][]{Ahn2012}.  After correcting for the host-galaxy
redshift, we measure integrated emission-line fluxes using standard
procedures via {\it splot} in IRAF, and follow Perez-Montero \& Diaz
(2003) to compute statistical errors\footnote{The measured H$\gamma$, H$\beta$, H$\alpha$, [O~III] $\lambda$4959,[O~III] $\lambda$5007, and [N~II] $\lambda$6584 emission-line fluxes are 
$34.1\pm3.3$, $88.5\pm3.6$, $290.9\pm3.4$, $81.5\pm3.2$, $232.7\pm4.8$, and $10.3\pm1.0 \times 10^{-17}$\,erg\,s$^{-1}$\,cm$^{-2}$ before extinction correction, respectively. 
Errors include statistical measurement uncertainties}. 
Using the SDSS Petrosian
magnitudes, correcting for Galactic and host-galaxy extinction (see
below), and applying $K$-corrections via {\it kcorrect} (v4.2, Blanton
\& Roweis 2007) at the host-galaxy redshift, we derive an absolute
magnitude of $M_r=-14.8\,$mag. This is an extremely underluminous dwarf
galaxy, well below the luminosities of the Large and Small Magellanic Clouds. It is one of
the least luminous host galaxies of CCSNe discovered by PTF, and it is not a
common host galaxy for a SN~Ic, even considering those 
found via untargeted surveys \citep{Arcavi2010}, except for SNe Ic-BL. 
In addition, the clear detections of numerous
bright emission lines allow us to apply standard diagnostics of the
star-formation rate (SFR), extinction, and metallicity to this galaxy.

\subsection{Star-Formation Properties}

Using the observed intensity ratio of H$\alpha$/H$\beta$, and assuming
the Case~B recombination value of 2.86 and the Cardelli et al. (1989)
extinction law with $R_V=3.1$, we estimate a reddening of
$E(B-V)=0.14\pm 0.05$\,mag. The values of the intensity ratios
[N~II]/H$\alpha$ and [O~III]/H$\beta$ indicate that the prominent
emission lines are due to recent vigorous star formation rather than
to an active galactic nucleus \citep{Baldwin1981}.  Furthermore, we do
not detect any absorption features nor a Balmer break that may
indicate the presence of an older stellar population.

From the measured rest-frame equivalent width of H$\beta$, ${\rm EW}_{{\rm H}\beta}=23.3\pm1\,$\AA, 
we derive an age of $6.4\pm0.1\,$Myr for the young stellar population using the relations of Levesque et al. (2010, \S 3.4 and Table 4)
and the measured metallicity of the host (for $Z=0.004$; see below).  
From the measured integrated H$\alpha$ emission-line flux, corrected for
extinction, we obtain $L({\rm H}\alpha)=(16\pm2) \times
10^{38}$\,erg\,s$^{-1}$, which translates to SFR(H$\alpha$)
$=0.013\pm0.002$\,M$_\odot$\,yr$^{-1}$ using the conversion from
Kennicutt (1998). Since the SDSS spectrum was obtained through a
$3\arcsec$ diameter fiber encompassing an area of $0.55\,$kpc$^2$ at the 
host-galaxy redshift, we measure a normalized SFR of $0.023\pm0.003$\,M$_\odot$\,yr$^{-1}$\,kpc$^{-2}$.
Since the Petrosian radius (from DR9) is $3.4\arcsec$, the fraction of global host-galaxy light entering the fiber,
which is also at the SN position, is $20\%$. 

\subsection{Oxygen Abundance}

For computing the oxygen abundance (Modjaz et al. 2011, and references therein), we correct
the detected emission-line fluxes of [O~III], [N~II], H$\beta$, and
H$\alpha$ for reddening, and employ the scales of Pettini \& Pagel
(2004, PP04-O3N2) and of Kewley \& Dopita (2002, KD02) to obtain
values of $12+{\rm log(O/H)}_{\rm PP04-O3N2}=8.12^{+0.04}_{-0.02}$ and
$12+{\rm log(O/H)}_{\rm KD02}=8.13^{+0.05}_{-0.04}$, respectively.  We
conclude that the metallicity of the host galaxy is
0.2--0.3\,Z$_\odot$, having used the solar oxygen abundance of
$12+{\rm log(O/H)}=8.69$ \citep{Asplund2009}.

The oxygen abundance of the PTF12gzk host is well below that of the
hosts of normal SNe~Ic found via untargeted surveys as presented by
Modjaz et al. (2011; mean $12+{\rm log(O/H)}_{\rm PP04}=8.7\pm0.1$)
and Sanders et al. (2012, $12+{\rm log(O/H)}_{\rm PP04}=8.61\pm0.2$). 
Indeed, it is much closer to that of the hosts of SNe~Ic-BL and
GRB-SNe \citep{Modjaz2008, Sanders2012}, and of SLSNe-I
\citep{Young2010,Stoll2011}.

\section{Discussion}

PTF12gzk is a luminous SN~Ic, at the high end of the SN~Ic luminosity
distribution \citep{Drout2011}. It exhibits a slow
rise of 18 days to its peak $r$-band magnitude, with $B$
peaking $\sim10$ days earlier. This is a large gap relative
to other SN~I, though similar to SN~2004aw \citep{Taubenberger2006};
it is caused by metal-line absorption from heavy elements in the outer
layers of the ejecta, as is evident from spectra taken after August 1.

A least-squares fit to a $f(t)\propto(t-t_0)^n$ behavior of our
well-sampled early photometry places the explosion date between 10
and 40\,hr prior to our discovery at the 95\% confidence level\footnote{The fireball models can at best
give an underestimate on the explosion date, since they do not incorporate the photon diffusion time.}. 
We cannot rule out the popular quadratic fireball model, but 
higher values of $n$ are possible as well for larger areas in the fit parameter space
(Figure 1, bottom panels).

Spectroscopically, PTF12gzk exhibits high expansion velocities,
$\sim30,000$\,km\,s$^{-1}$ (Si~II absorption velocity). 
Other SNe~Ic with similar velocities are broad-lined SNe~Ic (Figure 4), 
some of which are associated with GRBs (Woosley \& Bloom 2006, and references within), 
while no such association was determined for PTF12gzk (see also SN 2009bb; Soderberg et al. 2009).
Most similar is SN 2003lw, a SN associated with a GRB \citep{Mazzali2006a}. 
\begin{figure}[ht]
\centering
\scalebox{0.75}{\includegraphics{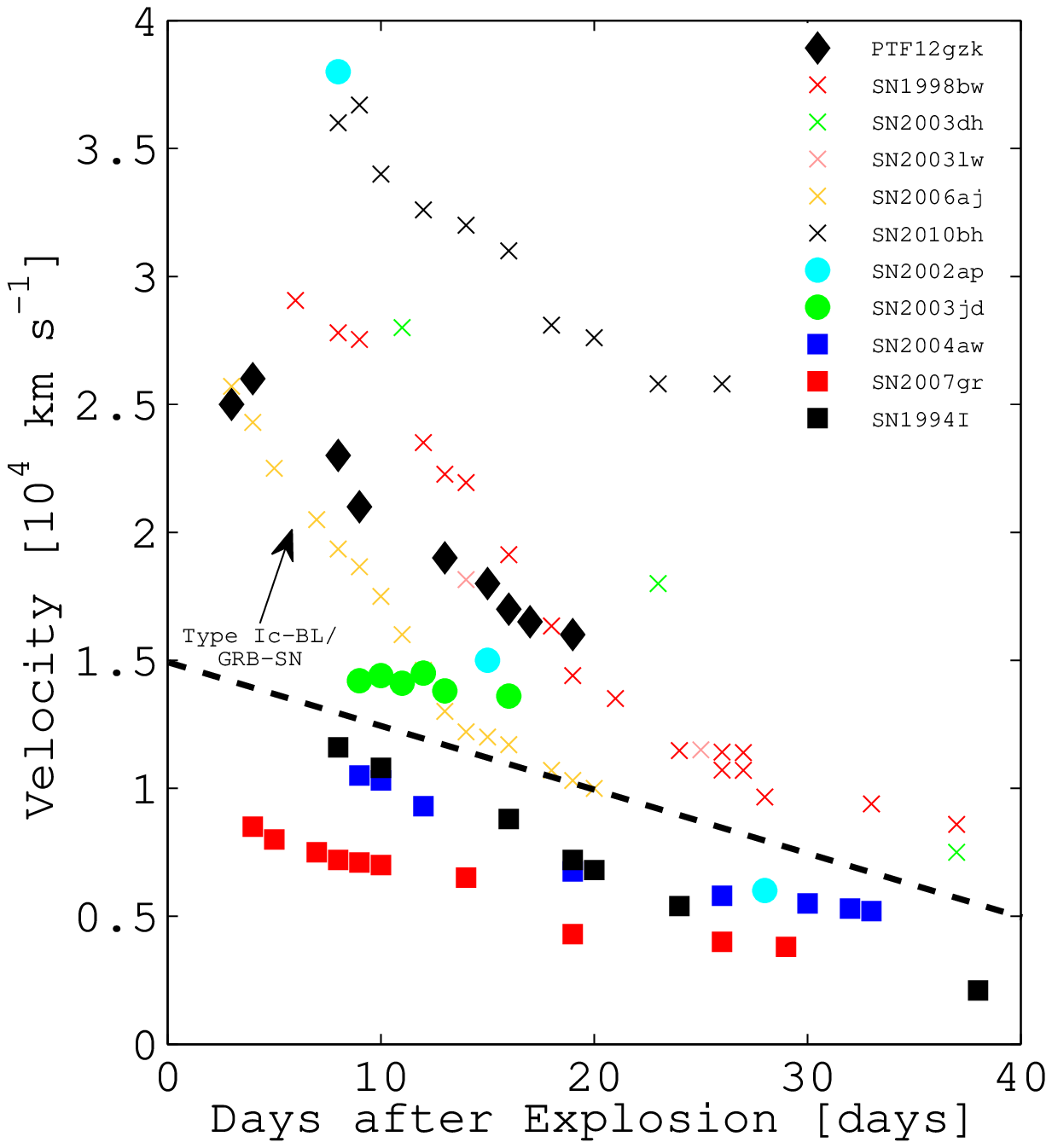}}
\caption{ {\scriptsize PTF12gzk has characteristic velocities of a broad-lined SN~Ic. All SNe above
 the dashed line, besides PTF12gzk, are  GRB-SNe (Xs) or Type Ic-BL 
 with no GRB association (circles), while those 
 below it are normal SNe~Ic (squares).
 Velocities are obtained through 
 modeling of the spectrum or through direct measurements of the Si~II 6355 \,\AA\ line
 (SNe 2010bh, 2002ap, 2003jd, 2004aw, and 2007gr).
} }
\end{figure}
A possible explanation is a burst misaligned with our line of sight, or a failed GRB. Such a
scenario is further supported by the host-galaxy characteristics,
resembling those of a broad-lined SN~Ic host galaxy. We know of no
typical SN~Ic exploding in a host with similar luminosity and oxygen
abundance.  The observed relatively narrow lines give a dispersion of
$\Delta v/v \approx 0.25$, compared to $\sim1$ in the case of
broad-lined SNe~Ic, and may suggest a nonspherical explosion geometry
\citep{Leonard2006}, or that the ejecta mass is high or has a very steep density gradient 
(a discussion on the effects of a steep density gradient on the LC can be found in Piro \& Nakar 2012). 
Late-time, nebular spectra will probe the geometry of the explosion in more detail.

From the Si~II line velocity at peak brightness of PTF12gzk
(15,300\,km\,s$^{-1}$ from the August 12 spectrum) and SN 2004aw
(12,400\,km\,s$^{-1}$; Deng et al., in prep.), and the rise time of
these two SNe, we use the following scaling relations (Arnett 1982; Mazzali et al. 2009;
see also Mazzali et al., in prep.) to estimate the physical properties
of PTF12gzk:
$\tau\approx\kappa^{1/2}\mathrm{M}^{3/4}\mathrm{E}^{-1/4}$ and
$v=(2E/M)^{1/2}$, where $\tau$ is the light-curve rise time, $E$ is
the kinetic energy, and $\kappa$ is the opacity.
The derived ejecta mass is 7.5\,M$_\odot$ (6--12\,M$_\odot$), pointing to a 
large initial progenitor mass of 25--35\,M$_\odot$, 
though the latter values are highly uncertain \citep{Mazzali2000}. We derive a
kinetic energy of $7.5 \times 10^{51}$\,erg [(5--10) $\times10^{51}$\,
erg]. Using $L_{\rm max}$ and $t_{\rm max}$ we get an estimated $^{56}$Ni mass of 0.37\,M$_\odot$ (scaling the PTF12gzk light curve to that of SN 2003dh; Mazzali et al. 2003).  Using the
$V$-band peak magnitude vs. nickel mass relation presented by
Perets et al. (2010), we derive a $^{56}$Ni mass of 0.35\,M$_\odot$, in
agreement with the results derived from the scaling relations.  These
physical properties, as well as the high expansion velocities and the
host galaxy, are unlike those of normal SNe~Ic, which
typically occur in large hosts and have low ejecta masses, and kinetic
energies, (1.7\,M$_\odot$, and $10^{51}$\,erg Drout et al. 2011. 
For nickel mass in Type Ic SNe see for example Taubenberger et al. 2006, and Sauer et al. 2006).
Instead, they are reminiscent of GRB-SNe \citep{Mazzali2009}.

PTF12gzk is a remarkable example of a SN~Ic in terms of expansion
velocities, evolution timescale, the ejected mass, and the kinetic
energy released in the explosion. We conclude that these properties
point to the explosion of a massive star deficient in H and He, at the
higher-mass end of SN~Ic progenitors. This further illustrates the
peculiar population of SNe~Ic exploding in dwarf hosts
\citep{Arcavi2010}, as seen also in the case of GRB-SNe and most
SLSNe-I.

PTF12gzk demonstrates the advantages of using an untargeted sky survey
such as PTF with an extensive network of instruments and telescopes in
various wavebands to detect and rapidly characterize unusual cases of
cosmic explosions.

%\section*{Acknowledgments:}
\bigskip
\smallskip

S.B. is supported by a Ramon Fellowship from ISA.
A.G. acknowledges support by grants from the ISF, BSF, GIF,
Minerva and the EU FP7/ERC.
A.V.F. and his group benefit from financial assistance from Gary \&
Cynthia Bengier, the Richard \& Rhoda Goldman Fund, the Sylvia \& Jim
Katzman Foundation, the Christopher R. Redlich Fund, the TABASGO
Foundation, NSF grants AST-0908886 and AST-1211916, and NASA/{\it HST} grant
GO-12530 from STScI (which is operated
by the AURA, Inc., under NASA contract NAS 05-26555).
P.A.M. and E.P. acknowledge financial support from grants INAF PRIN
2011 and ASI/INAF I/088/06/0.
M.I. and Y.J. were supported by the Creative Initiative program of the NRFK.
M.M.K. acknowledges Hubble and Carnegie-Princeton Fellowships.
D.C.L. is supported by NSF grant AST-1009571.
E.O.O. acknowledges the Arye Dissentshik career development chair and
a grant from the Israeli MOST.

PTF is a collaboration of Caltech, LCOGT, the Weizmann Institute, LBNL/NERSC,
Oxford, Columbia, IPAC, and UC Berkeley.  
The Liverpool Telescope is operated on the island of La Palma by
Liverpool John Moores University in the Spanish Observatorio del Roque
de los Muchachos  support from the UK STFC.
Construction of the LAIWO camera was supported by
the MPIA, GIF, and the ISF. 
We are grateful for the assistance of the staff at the various
observatories used to obtain data.
This research has made use of the NASA/IPAC
Extragalactic Database (NED) which is operated by JPL, 
Caltech, under contract with
NASA.

\end{document}